\documentclass{aa}  

\usepackage{graphicx}
\usepackage{amsmath}
\usepackage{amssymb}
\usepackage{mathtools}
\usepackage{bm}
\usepackage{txfonts}
\begin{document}

   \title{The Effects of Magnetic Field Strength on Properties of Wind Generated from Hot Accretion Flow}
%   \subtitle{I. Overviewing the $\kappa$-mechanism}

   \author{De-Fu Bu
          \inst{1}
          \and
          Amin Mosallanezhad 
          \inst{2} 
              }

   \institute{Key Laboratory for Research in Galaxies and Cosmology, Shanghai Astronomical Observatory, Chinese Academy of Sciences, 80 Nandan Road, Shanghai 200030, China \\
              \email{dfbu@shao.ac.cn}
         \and
             Key Laboratory for Research in Galaxies and Cosmology, Department of Astronomy, University of Science and Technology of China, Hefei, Anhui 230036, China \\
             \email{mosallanezhad@ustc.edu.cn}
             }

\titlerunning{Wind from hot accretion flow}
\authorrunning{Bu \& Mosallanezhad}
%   \date{Received September 15, 1996; accepted March 16, 1997}

% \abstract{}{}{}{}{} 
% 5 {} token are mandatory
 
  \abstract
   {Observations indicate that wind can be generated in hot accretion flow. By performing numerical simulations, Yuan et al. studied the detailed properties of wind generated from weakly magnetized accretion flow. However, properties of wind generated from strongly magnetized hot accretion flow have not been studied.}
   {In this paper, we study the properties of wind generated from both weakly and strongly magnetized accretion flow. We focus on how the magnetic field strength affects the wind properties.}
   {We solve time-steady two-dimensional magnetohydrodynamic (MHD) equations of black hole accretion in the presence of large-scale magnetic field. We assume self-similarity in radial direction. The magnetic field is assumed to be evenly symmetric with the equatorial plane.}
   {We find that wind exists in both weakly and strongly magnetized accretion flow. When magnetic field is weak (magnetic pressure is more than $ 2 $ orders of magnitude smaller than gas pressure), wind is driven by gas pressure gradient and centrifugal forces. When magnetic field is strong (magnetic pressure is slightly smaller than gas pressure), wind is driven by gas pressure gradient and magnetic pressure gradient forces. The power of wind in strongly magnetized case is just slightly larger than that in weakly magnetized case. The power of wind lies in a range $ P_W \sim 10^{-4}-10^{-3} \dot M_{\rm in} c^2 $, with $ \dot M_{\rm in} $ and $ c $ being mass inflow rate and speed of light, respectively. The possible role of wind in active galactic nuclei feedback is briefly discussed.}
   {}


   \maketitle
%
%-------------------------------------------------------------------

\section{Introduction} \label{sec:intro}

Advection-dominated accretion flow (ADAF) can have either low or high accretion rate. When the accretion rate is much larger than the Eddington rate, due to large optical depth, the photons are trapped in the flow and advected to the black hole. The high accretion rate ADAF is called slim disk (\citealt{Abramowicz et al. 1988}). When the accretion rate is significantly smaller than the Eddington rate, the flow density is very low. In this case, the radiation is not important. The flow temperature is close to the Virial temperature. The low accretion rate ADAF is called hot accretion flow.

Hot accretion flow model became popular since the 1990s (\citet{Narayan and Yi 1994,Narayan and Yi 1995}; \citealt{Abramowicz et al. 1995}; \citet{Kato et al. 1998}; \citealt{Narayan et al. 1998}). In low-luminosity active galactic nuclei (LLAGNs), the black hole is believed to accrete matter in hot accretion flow mode (e.g. \citealt{Ho 2008}; \citealt{Antonucci 2012}; \citealt{Done 2014}). In hard/quiescent states of black hole X-ray binaries, the black hole also accretes matter in hot accretion flow mode (e.g. \citealt{Esin et al. 1997}; \citealt{Fender et al. 2004}; \citealt{Zdziarski and Gierlinski 2004}; \citealt{Narayan 2005}; \citealt{Remillard and McClintock 2006}; \citealt{Narayan and McClintock 2008}; \citealt{Belloni 2010}; \citealt{Wu et al. 2013}; \citealt{Yuan and Narayan 2014}). In recent years, lots of numerical simulations have been performed to study properties of hot accretion flow (e.g. \citealt{Stone et al.1999}; \citealt{Igumenshchev and Abramowicz 1999, Igumenshchev and Abramowicz 2000}; \citealt{Hawley et al. 2001}; \citealt{Machida et al. 2001}; \citealt{Stone and Pringle 2001}; \citealt{Hawley and Balbus 2002}; \citealt{De Villiers et al. 2003}; \citealt{Pen et al. 2003}; \citealt{Beckwith et al. 2008}; \citealt{Pang et al. 2011}; \citealt{Tchekhovskoy et al. 2011}; \citealt{Tchekhovskoy and McKinney 2012}; \citealt{Yuan et al. 2012a, Yuan et al. 2012b}; \citealt{McKinney et al. 2012}; \citealt{Narayan et al. 2012}; \citealt{Li et al. 2013}; \citealt{Sadowski et al. 2013}; \citealt{Moscibrodzka et al. 2014}).

In recent years, there are observations of both LLAGNs (e.g. \citealt{Crenshaw and Kraemer 2012}; \citealt{Tombesi et al. 2010a, Tombesi et al. 2014}; \citealt{Wang et al. 2013}; \citealt{Cheung et al. 2016}) and the hard state of black hole X-ray binaries (\citealt{Homan et al. 2016}). These observations indicate that winds can be generated from hot accretion flow. The origin of wind in weakly magnetized hot accretion flow has been studied by numerical simulation (\citealt{Yuan et al. 2015}). In \citealt{Yuan et al. 2015}, magnetic pressure is smaller than gas pressure by a factor of several tens. It is found that wind is driven by the combination of gas pressure gradient, magnetic pressure gradient and centrifugal forces. The origin of wind has also been studied by analytical works. In these works, the effects of magnetic field (Lorentz force) are neglected (e.g. \citealt{Begelman 2012}; \citealt{Gu 2015}).

It is very important to study wind properties. The reasons are as follows. First, wind is an important ingredient of accretion physics. If wind is present, the mass accretion rate can no longer be a constant with radius.  The change of mass accretion rate profile changes the gas density profile.The change of density profile can change spectrum of a black hole accretion system (e.g. \citealt{Quataert and Narayan 1999}; \citealt{Yuan et al. 2003}). Second, there is consensus that AGNs feedback can affect the evolution of their host galaxies significantly (e.g. \citealt{Ciotti and Ostriker 1997, Ciotti and Ostriker 2001, Ciotti and Ostriker 2007}; \citealt{Proga et al. 2000}; \citealt{Novak et al. 2011}; \citealt{Gan et al. 2014, Gan et al. 2017}; \citealt{King and Pounds 2015}; \citealt{Ciotti et al. 2017}). Wind can push away the gas at sub-parsec and parsec scales. Consequently, the feeding rate of the central black hole can be significantly affected. The change of gas properties at parsec scale can also affect the star formation rate (e.g. \citealt{Ciotti et al. 2017}). Before studying the feedback by AGN wind, we should study the properties of wind.

Both cold disk (\citealt{Shakura and Sunyaev 1973}) and hot accretion flow can produce wind. Luminous AGNs are powered by standard thin disk (\citealt{Shakura and Sunyaev 1973}). Blueshifted absorption lines of highly ionized iron are frequently observed in luminous AGNs. The large blueshift of lines indicates that wind has velocity of $ 0.1-0.3\ c $, with $c$ being speed of light. Such high velocity wind is called ultra-fast outflows (UFOs, \citet{Tombesi et al. 2010b, Tombesi et al. 2011}). UFOs are generated inside $ 100 $ Schwarzschild radius. The power of UFOs can be $ 5/1000\ L_\text{Edd} $, with $ L_\text{Edd} $ being Eddington luminosity (\citealt{Nomura and Ohsuga 2017}).  The status of study of wind from hot accretion flows is quite different from the case of a cold accretion disk. In this case, the observational data is much fewer compared to the case of cold disk. This is mainly because the gas in the wind from a hot accretion flow is very hot thus generally fully ionized. So it is very difficult to detect them by the usual absorption-line spectroscopy. But still, in recent years, we have gradually accumulated more and more observational evidences for wind from low-luminosity sources in which we believe a hot accretion flow is operating (e.g. \citealt{Crenshaw and Kraemer 2012}; \citealt{Tombesi et al. 2010a, Tombesi et al. 2014}; \citealt{Wang et al. 2013}; \citealt{Cheung et al. 2016}; {Homan et al. 2016} ). The properties of wind from hot accretion flow are mainly studied by numerical simulations. Simulation works find that poloidal velocity of wind is approximately $ 0.2\ v_k $, with $ v_k $ being the Keplerian velocity of the location where wind is generated.The power of wind is $ 1/1000\ \dot{M} c^2 $, with $ \dot{M} $ being black hole accretion rate (\citealt{Yuan et al. 2015}). 

We define plasma $ \beta = p/p_{\rm mag} $, where $ p $ and $ p_{\rm mag} $ are the gas pressure and the magnetic pressure, respectively. Numerical simulations found that hot accretion flow can have either large or small $ \beta $.  When $ \beta $ is large ($ \beta \gtrsim 10 $), magnetic pressure is much smaller than gas pressure, accretion flow is weakly magnetized. In this case, angular momentum is transferred by Maxwell stress associated with MHD turbulence driven by magneto-rotational instability (MRI; \citealt{Balbus and Hawley 1998}). The weakly magnetized hot accretion flow is called `standard and normal evolution' (SANE) model (\citealt{Narayan et al. 2012}). When $ \beta $ is small (e.g. $ \beta \lesssim 10 $), the magnetic pressure becomes comparable to the gas pressure. In this case, the magnetic field is too strong that MHD turbulence is suppressed. Angular momentum is transferred by strong ordered large scale magnetic field (\citealt{Stone and Norman 1994}; \citealt{Narayan et al. 2012}; \citealt{McKinney et al. 2012}; \citealt{Tchekhovskoy and McKinney 2012}). The strongly magnetized accretion flow is called `magnetically arrested disc' (MAD; e.g. \citealt{Narayan et al. 2012}). For the MAD, the accretion rate is strongly suppressed by the strong magnetic field.

The detailed properties of the wind in SANE model have been investigated by \citealt{Yuan et al. 2012b} (see also \citealt{Yuan et al. 2015}). However, properties of wind in strongly magnetized hot accretion flow have not been studied. In \citealt{Yuan et al. 2015}, it is found that the wind is driven by the combination of magnetic pressure gradient, gas pressure gradient and centrifugal forces. Therefore, the strength of magnetic field should be an important factor in determining properties of wind. In this paper, we study how the wind properties change with the changing of magnetic field strength.We study both weakly and strongly magnetized hot accretion flow. We obtain the two-dimensional axisymmetric steady solution of hot accretion flow in the presence of magnetic
field. We assume radial self-similarity as in many previous works.

Analytical works assuming radial self-similarity have been done by many authors. In one-dimensional solution, wind is usually assumed to be present (e.g. \citealt{Blandford and Begelman 1999}; \citealt{Akizuki and Fukue 2006}; \citealt{Abbassi et al. 2008}; \citealt{Zhang and Dai 2008}; \citealt{Bu et al. 2009}). Wind has been automatically found in two-dimensional solutions (e.g. \citealt{Xu and Chen 1997}; \citealt{Blandford and Begelman 2004}; \citealt{Xue and Wang 2005}; \citealt{Tanaka and Menou 2006}; \citealt{Jiao and Wu 2011}; \citealt{Mosallanezhad et al. 2014}; \citealt{Gu 2015}; \citealt{Samadi and Abbassi 2016}). However, we note here that in all above mentioned works, magnetic field is not taken into account. It is well known that magnetic field plays a significant role in the dynamics of accretion flow. Therefore, it is necessary to obtain the accretion solutions in the presence of magnetic field. In this paper, we will study how do the wind properties change with the changing of magnetic field strength.

This paper is organized as follows. In section \ref{sec:equations}, we introduce the basic MHD equations and assumptions. The detailed descriptions of numerical solutions are described in section \ref{sec:results}. In section \ref{sec:summery}, we provide a summary and briefly discuss the implications of our results.

%--------------------------------------------------------------------
\section{Basic Equations} \label{sec:equations}

%-------------------------------------- Two column figure (place early!)
%   \begin{figure*}
%   \centering
%   %%%\includegraphics{empty.eps}
%   %%%\includegraphics{empty.eps}
%   %%%\includegraphics{empty.eps}
%   \caption{Adiabatic exponent $\Gamma_1$.
%               $\Gamma_1$ is plotted as a function of
%               $\lg$ internal energy $\mathrm{[erg\,g^{-1}]}$ and $\lg$
%               density $\mathrm{[g\,cm^{-3}]}$.}
%              \label{FigGam}%
%    \end{figure*}
%

In this section, we derive the basic equations for steady-state, axisymmetric hot accretion flows incorporating magnetic fields. We adopt spherical coordinates $ (r, \theta, \phi) $. General-relativistic effects are neglected and we use Newtonian potential $ \psi = - GM/r $, where $ G $ is the gravitational constant and $ M $ is the mass of the central black hole. The basic resistive MHD equations of hot accretion flow can be described as:

\begin{equation} \label{Continuity}
	\frac{\partial \rho}{\partial t} + \bm{\nabla} \cdot \left( \rho \bm{v} \right) = 0,
\end{equation}
\begin{equation} \label{Momentum}
	\rho  \left[ \frac{\partial \bm{v}}{\partial t}  + \left(  \bm{v} \cdot \bm{\nabla} \right)  \bm{v}  \right] = - \rho \bm{\nabla} \psi -   \bm{\nabla} p + \bm{\nabla} \cdot \bm{T} + \frac{\bm {J}\times \bm{B}}{c} ,
\end{equation}
\begin{equation} \label{Energy}
	\frac{\partial e}{\partial t} + \bm{\nabla} \cdot \left( e \bm{v} \right) + p \bm{\nabla} \cdot \bm{v} = q^{+} - q_{\rm rad}^{-} \equiv f q^{+},
\end{equation}
\begin{equation} \label{Induction}
	\frac{\partial \bm{B}}{\partial t} = \bm{\nabla} \times \left(  \bm{v} \times \bm{B} - \frac{4 \pi}{c} \eta \bm{J} \right),
\end{equation}
\begin{equation} \label{DelB}
	\bm{\nabla} \cdot \bm{B} = 0,
\end{equation}
 where, $ \rho $ is the density, $ \bm{v} = (v_{r}, v_{\theta}, v_{\phi}) $ is the velocity, $ p $ is the gas pressure, $ \bm{T} $ is the viscous stress tensor, $ \bm{J} = (c/4\pi) \bm{\nabla} \times \bm{B} $ is the current density, $ \bm{B} = (B_{r}, B_{\theta}, B_{\phi}) $ is the magnetic field and, $ e $ is the internal energy density of the gas. We adopt an adiabatic equation of state, $ p = ( \gamma - 1) e $, where $ \gamma $ is the adiabatic index of the gas which is set to $ 5/3 $ here. In the energy equation (\ref{Energy}), $ q^{+} $ is the heating rate, $ q^{-}_{\rm rad} $ is the radiative cooling rate and, $ f $ represents the advection factor describing the fraction of the heating energy stored in the gas and advected towards the central black hole. In the induction equation (\ref{Induction}), $ \eta = c^2/(4 \pi \sigma_{\rm e}) $ is the magnetic diffusivity, where $ \sigma_{\rm e}$ denotes the electric conductivity. Note here that, we include the dissipation term in Equation (\ref{Induction}) in order to obtain a steady solution.  If the dissipation term is neglected, there will be no steady state solution due to continuous accumulation of magnetic flux.
 
Numerical simulation found that the azimuthal component of viscosity is much larger than other components (\citealt{Stone and Pringle 2001}). Therefore in this study, we set the viscous tensor to have only the azimuthal component, i.e.,
\begin{equation} \label{Trp}
    T_{r \phi} = \mu r \frac{\partial}{\partial r} \left( \frac{v_{\phi}}{r} \right),
\end{equation}
where  $ \mu(\equiv \rho \nu) $ is the dynamical viscosity coefficient and, $ \nu $ is the kinematic viscosity. We model viscosity with the standard $ \alpha $-prescription (\citealt{Shakura and Sunyaev 1973}), where the kinematic coefficient of the viscosity $ \nu $ can be expressed as
\begin{equation} \label{nu}
	\nu = \frac{\alpha p}{\rho \Omega_{\rm K}}.
\end{equation}

Here, $ \Omega_{\rm K} = (GM/r^{3})^{1/2}  $ is the Keplerian angular velocity and $ \alpha $ is a constant viscosity parameter. We assume both viscosity and magnetic dissipation can heat the gas. Therefore, the heating rate $ q^+ $ will be decomposed into two terms,
\begin{equation} \label{heating rate}
	q^{+} = q_{\rm vis} + q_{\rm res},
\end{equation}
with
\begin{equation} \label{q_vis}
	q_{\rm vis} = T_{r\phi} r \frac{\partial}{\partial r} \left(\frac{v_{\phi}}{r} \right),
\end{equation}
\begin{equation} \label{q_res}
	q_{\rm res} = \frac{4\pi}{c^2} \eta \bm{J}^2.
\end{equation}
where, $ q_{\rm vis} $, and $ q_{\rm res} $ are the viscous heating and the magnetic field dissipation heating, respectively. We set the magnetic diffusivity to be $ \eta = \eta_{0} p/ (\rho \Omega_{\rm K}) $ to satisfy the radially self-similar assumption.

Global and local numerical MHD simulations of black hole accretion disks found that magnetic field can be decomposed into an ordered large-scale component and a turbulent component (e.g. \citealt{Machida et al.2006}; \citealt{Johansen and Levin 2008}; \citealt{Bai and Stone 2013}; \citealt{Zhu and Stone 2017}). Both large-scale and turbulent components can transfer angular momentum. The dissipation of magnetic field can heat the gas. In this paper, the magnetic field $ \bm{B} $ in Equations (\ref{Momentum}), (\ref{Induction}) and (\ref{DelB}) represents the large-scale ordered component. The term $ q_{\rm res} $ represents the heating by dissipation of large scale component of magnetic field. The angular momentum transfer by small-scale turbulent magnetic field is modeled by the viscous force $ \nabla \cdot \mathbf{T} $ in equation (\ref{Momentum}). The dissipation heating of gas by small scale turbulent magnetic field is modeled by the viscous heating $ q_{\rm vis} $.

The magnetic field configuration is assumed to be evenly symmetric about the equatorial plane. This kind of symmetry is immensely implemented by previous works to study MHD wind (e.g. \citealt{Blandford and Payne 1982}; \citealt{Lovelace et al. 1994}; \citealt{Cao 2011}; \citealt{Li and Begelman 2014}).
\begin{gather}
  B_{r}(r, \theta) = - B_{r}(r, \pi - \theta), \label{even symmetry r} \\
  B_{\theta}(r, \theta) = + B_{\theta}(r, \pi - \theta), \label{even symmetry th}\\
  B_{\phi}(r, \theta) = - B_{\phi}(r, \pi - \theta), \label{even symmetry phi}
\end{gather}
Radial component of magnetic field can be stretched into azimuthal component due to shear of accretion flow. Thus, the radial and azimuthal components of magnetic field have opposite sign. In this paper, we study the region above the equatorial plane. Radial component of magnetic field is set to be positive ($ B_r > 0 $). Accordingly, azimuthal component of magnetic field is negative ($ B_{\phi} < 0 $).

We assume time-steady and axisymmetric flow ($ \partial/\partial t = \partial/\partial \phi = 0 $), and adopt self-similar approximation to remove radial dependence of physical quantities. In terms of a fiducial radial distance, $ r_{0} $, the self-similar solutions are defined as power-law form of $ r/r_{0} $. Therefore, the MHD equations of the flow admit the following radial scaling relations for the variables,

\begin{equation} \label{self_vs}
	\bm{v} (r, \theta) = \sqrt{\frac{GM}{r_{0}}} \left( \frac{r}{r_{0}} \right)^{-1/2} \bm{v}(\theta),
\end{equation}
\begin{equation} \label{self_rho}
	\rho(r, \theta) = \frac{M}{r_{0}^{3}} \left( \frac{r}{r_{0}} \right)^{-n} \rho(\theta),
\end{equation}
\begin{equation} \label{self_p}
	p(r, \theta) = \frac{GM^{2}}{r_{0}^{4}} \left( \frac{r}{r_{0}} \right)^{-n-1} p(\theta),
\end{equation}
\begin{equation} \label{self_bs}
	\bm{B} (r, \theta) = \sqrt{\frac{GM^{2}}{r_{0}^{4}}} \left( \frac{r}{r_{0}} \right)^{-(n/2)-(1/2)} \bm{b}(\theta).
\end{equation}

Numerical simulations of hot accretion flow show that the radial profile of density can be described as a power law function of $ r $ as $ \rho \propto r^{-n} $, with $ 0.5 < n < 1 $ (e.g. \citealt{Stone et al.1999}; \citealt{Yuan et al. 2012a}). Therefore, in this paper we assume that $ \rho \propto r^{-n} $. According to the power law function of density, the power-law function of magnetic field and gas pressure are set to satisfy the radially self-similar condition. In this paper, we set $ n=0.85 $. We have done some tests with other values of $ n $. We found that if $ n $ is slightly changed, the results are not changed much.

Under the above mentioned assumptions, Equations (\ref{Continuity})-(\ref{DelB}) can be simplified as

\begin{equation} \label{con}
	\rho  \left[  \left( \frac{3}{2} - n \right) v_{r} + v_{\theta} \cot \theta + \frac{dv_{\theta}}{d \theta} \right] + v_{\theta} \frac{d \rho}{d \theta} = 0,
\end{equation}
\begin{multline} \label{mom_r}
	\rho \left[ -\frac{1}{2} v_{r}^{2} + v_{\theta} \frac{dv_{r}}{d \theta}  - v_{\theta}^{2} - v_{\phi}^{2} \right] =  - \rho + (n + 1) p \\ + \frac{1}{4 \pi} \left( j_{\theta} b_{\phi} - j_{\phi} b_{\theta} \right),
\end{multline}
\begin{multline} \label{mom_t}
	\rho \left[ \frac{1}{2} v_{r} v_{\theta} + v_{\theta} \frac{dv_{\theta}}{d \theta} - v_{\phi}^{2} \cot \theta \right] = - \frac{d p}{d \theta} \\ + \frac{1}{4 \pi} \left( j_{\phi} b_{r} - j_{r} b_{\phi} \right),
\end{multline}
\begin{multline} \label{mom_p}
	\rho \left[ \frac{1}{2} v_{r} v_{\phi} + v_{\theta} \frac{dv_{\phi}}{d \theta} + v_{\theta} v_{\phi} \cot \theta \right] =  \frac{3}{2}(n - 2) \alpha v_{\phi} p \\ + \frac{1}{4 \pi} \left( j_{r} b_{\theta} - j_{\theta} b_{r} \right),
\end{multline}
\begin{multline} \label{engy}
	\left( \frac{3}{2} \gamma - n - 1\right) v_{r} p + v_{\theta} \frac{d p}{d \theta} + \gamma p \left[ \frac{d v_{\theta}}{d \theta} + v_{\theta} \cot \theta \right] = \\ f (\gamma - 1) \left[ \frac{9}{4} \alpha p v_{\phi}^{2} + \frac{\eta}{4 \pi} \left( j_{r}^{2} + j_{\theta}^{2} + j_{\phi}^{2} \right) \right],
\end{multline}
\begin{multline} \label{ind_p}
	\frac{n}{2} \left( v_{r} b_{\phi} - v_{\phi} b_{r} \right) + v_{\phi} \frac{d b_{\theta}}{d \theta} + b_{\theta} \frac{d v_{\phi}}{d \theta} - v_{\theta} \frac{d b_{\phi}}{d \theta} - b_{\phi} \frac{d v_{\theta}}{d \theta} \\ + \frac{n}{2} \eta j_{\theta} + \eta \frac{d j_{r}}{d \theta} + j_{r} \frac{d \eta}{d \theta} = 0,
\end{multline}
\begin{equation} \label{delb}
	\frac{d b_{\theta}}{d \theta} - \frac{1}{2}( n - 3) b_{r} + b_{\theta} \cot \theta = 0,
\end{equation}
where,
\begin{equation} \label{jr}
	j_{r} = \frac{db_{\phi}}{d \theta} + b_{\phi} \cot \theta,
\end{equation}
\begin{equation} \label{jt}
	j_{\theta} = \frac{1}{2} (n - 1) b_{\phi},
\end{equation}
\begin{equation} \label{jp}
	j_{\phi} = - \frac{1}{2} (n - 1) b_{\theta} - \frac{db_{r}}{d \theta}.
\end{equation}

The above differential equations consist of eight variables: $ v_{r}(\theta) $, $ v_{\theta}(\theta) $, $ v_{\phi}(\theta) $, $ \rho(\theta) $, $ p(\theta) $, $ b_{r}(\theta) $, $ b_{\theta}(\theta) $ and $ b_{\phi}(\theta) $. By assuming even symmetry for the accretion flow about the equatorial plane, the boundary conditions require,
\begin{equation}\label{boundry conditions}
	\frac{d v_{r}}{d \theta} = \frac{d v_{\phi}}{d \theta} = \frac{d \rho}{d \theta} = \frac{d p}{d \theta} =  \frac{d b_{\theta}}{d \theta} = b_{r} = b_{\phi}=v_{\theta} = 0.
\end{equation}

%In principle, there should be a second boundary condition at the rotation axis to solve two-point boundary problem. Unfortunately because of some technical difficulties such as singularity at $ \theta = 0^{\circ} $, we could not obtain such solutions. Instead, we only require the solution satisfying the boundary condition at $ \theta = \pi/2 $. We think the solution obtained in this way should still be physically meaningful (see the explanation of this point in next section). To do so, 
We set the density at the equatorial plane $ \rho(\pi/2) = 1 $. Also, the magnetic field strength is set to,

\begin{equation} \label{beta_0}
	\beta_{0} =  \frac{ 8 \pi p(\pi/2)}{b^{2}_{\theta}(\pi/2)}
\end{equation}

%-----------------------------------------------------------------
\section{Results} \label{sec:results}

Equations (\ref{con})-(\ref{delb}) are solved numerically. We set $ B_r $ and $ B_\phi $ to be null at the equatorial plane. Above the equatorial plane, we set $B_r$ to be positive, $ B_\phi $ is generated by shear of the flow and should be negative. The MHD equations are integrated from $ \theta=\pi/2 $ towards the rotational axis ($ \theta=0 $). We find that $ B_\phi $ is negative at the beginning of integration. However, $ B_\phi $ becomes positive at an angle $ \theta_{\rm s} $ near the rotation axis. We stop integration at $ \theta_{\rm s} $. We believe the solution in the region of $ \theta_{\rm s} < \theta < \pi/2 $ is still physical \footnote{Two-dimensional results found in \citealt{Jiao and Wu 2011} are similar in the sense that they also had to stop in their integration at a certain $ \theta > 0 $.}. The reasons are as follows. First, the solutions satisfy boundary conditions at $ \theta = \pi/2 $. Second, the physical quantities at $ \theta_{\rm s} $ are physical. Therefore, we can reasonably treat these values as boundary conditions at $ \theta_{\rm s} $. In fact, in \citealt{Mosallanezhad et al. 2016}, we also study properties of accretion flow by assuming radial self-similarity. It is shown that the main properties of the solutions obtained in this way are in good agreement with those obtained in \citealt{Yuan et al. 2015} from numerical simulations.

\begin{figure*}
\begin{center}
\includegraphics[width=140mm, angle=0]{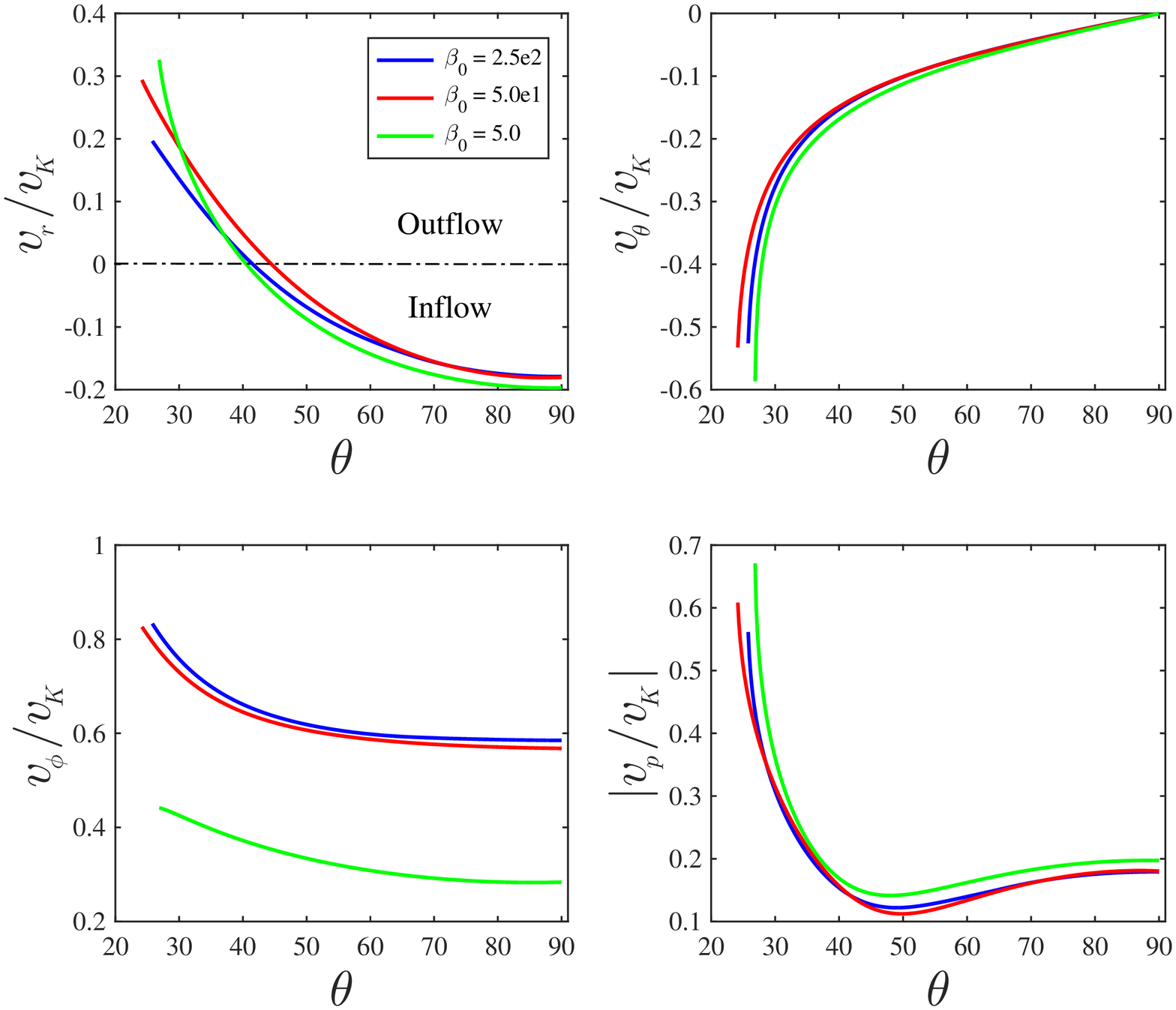}
\includegraphics[width=140mm, angle=0]{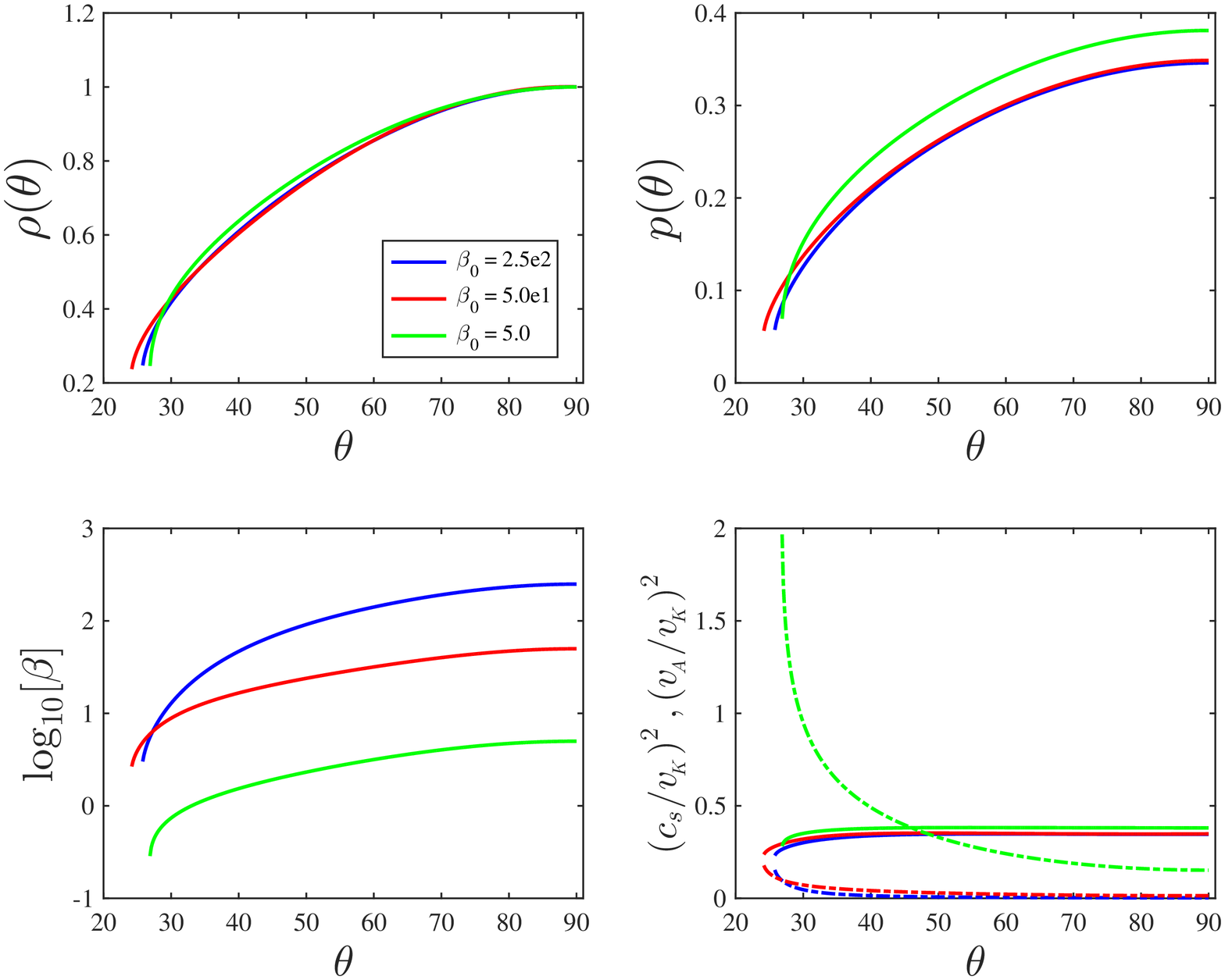}
\caption{Angular profiles of variables. The blue, red and green lines are for $\beta_0 = 250$, $ 50 $ and $ 5 $, respectively. $ \beta_0 $ is the ratio of gas pressure to magnetic pressure at the equatorial plane. Top-left panel plots radial velocity. Top-right panel plots $ v_\theta $. The left panel of the second row plots azimuthal velocity. Right panel of the second row plots poloidal velocity ($ v_p = \sqrt{v_{r}^2 + v_{\theta}^2} $). Left panel of the third row plots density. Right panel of the third row plots gas pressure. Bottom-left panel plots $ \beta $. In bottom-right panel, the solid lines are for sound speed, the dashed lines are for alfven speed. The top-left panel shows that the radial velocity changes its sign at $ \theta \sim 40^{\circ} $. \label{velocites}}
\end{center}
\end{figure*}

The parameters we adopt are  $ \alpha=\eta_{0}=0.1 $. Numerical simulations of hot accretion flow show that the radial profile of density can be described as a power law function of $ r $ as $ \rho \propto r^{-n} $, with $ 0.5 < n < 1 $ (e.g. \citealt{Stone et al.1999}; \citealt{Yuan et al. 2012a}). In this paper, we set $ n=0.85 $. We also assume that radiation is not important and set the advection factor $ f = 1 $. In this paper, we have three models. In model A, the plasma beta at the equatorial plane is $ \beta_{0} = 250 $. In models B and C, $ \beta_{0} = 50 $ and $ 5 $, respectively. The magnetic field is strongest in model C. The magnetic field strength in model B is moderate. In model A, the magnetic field is the weakest.

Figure \ref{velocites} plots the angular profiles of physical quantities. The blue, red and green lines are for $ \beta_0 = 250 $ (model A), $ 50 $ (model B) and $ 5 $ (model C), respectively.  It is clear that at the region close to the equatorial plane $ \theta > 40^{\circ} $, the radial velocity is negative, gas flows towards the black hole. The region $ \theta > 40^{\circ} $ is inflowing region. In the region $ \theta < 40^{\circ} $, the radial velocity is positive. The region $ \theta < 40^{\circ} $ is wind region. This result is fully consistent with that obtained by simulations (\citealt{Yuan et al. 2015}; see also \citealt{Narayan et al. 2012}; \citealt{Sadowski et al. 2013}). In those works, it is found that inflow is present around the equatorial plane while wind is present in the polar region.

The azimuthal velocity (see Figure \ref{velocites}) in the highly magnetized case ($ \beta_0=5 $) is significantly smaller than those in weakly magnetized cases ($ \beta_0=50 $ and $ 250 $). The reason is as follows. In highly magnetized accretion flow, Maxwell stress is large. Angular momentum can be transferred outward very efficiently. Therefore, specific angular momentum (or equivalently azimuthal velocity) of highly magnetized accretion flow is much smaller. Numerical simulations also find that the specific angular momentum of gas in MAD model is significantly smaller than that in SANE model (\citealt{Narayan et al. 2012}; \citealt{Tchekhovskoy and McKinney 2012}; \citealt{McKinney et al. 2012}).

\begin{figure*}
\begin{center}
\includegraphics[width=\textwidth]{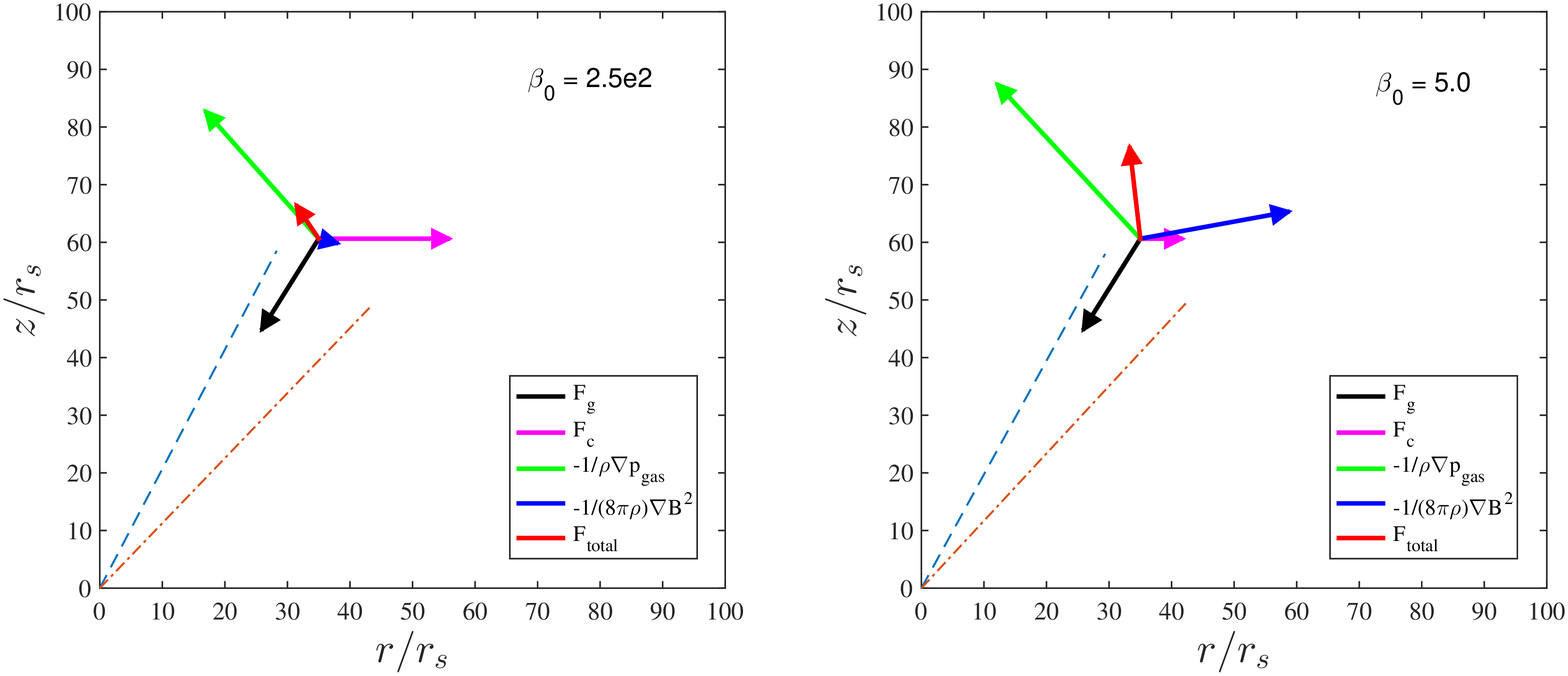}
\caption{The force analysis at the wind region to show the driving mechanism of wind. The left and right panels are for models A and C, respectively. The length of the arrows
schematically denotes the magnitude of the forces. $ F_g $: gravitational force; $ F_c $: centrifugal force; $ F_{\rm total} $: the total force. \label{schematic}}
\end{center}
\end{figure*}

The quantities $ v_{\theta} $ and $ v_{\phi} $ increase with decreasing $ \theta $. This is consistent with that found in \citealt{Jiao and Wu 2011}. In the wind region ($\theta < 40^\circ$), we find that the poloidal velocity ($ v_p=\sqrt{v_{r}^{2} + v_{\theta}^{2}} $) of wind in the strongly magnetized flow is larger than that in weakly magnetized flow (see Figure \ref{velocites}). In order to study the reason for the difference of poloidal velocity in different models, we calculate the forces at the wind region. Figure \ref{schematic} shows the result. The left and right panels are for models A and C, respectively. In the weakly magnetized case (model A), the dominant driving forces for wind are the centrifugal force and the gradient of the gas pressure. In the strongly magnetized case (model C), the centrifugal force is much smaller due to the low rotational velocity in this model (see left panel of the second row of Figure \ref{velocites}). The magnetic pressure gradient force in model C is much stronger than that in model A. The gas pressure gradient force is larger in model C than that in model A. The total force in model C is larger than that in model A. Therefore, the poloidal velocity of wind in model C is biggest. We note that the poloidal velocity of wind in model C is just slightly larger than those in models A and B.

\begin{figure*}
\begin{center}
\includegraphics[width=120mm, angle=0]{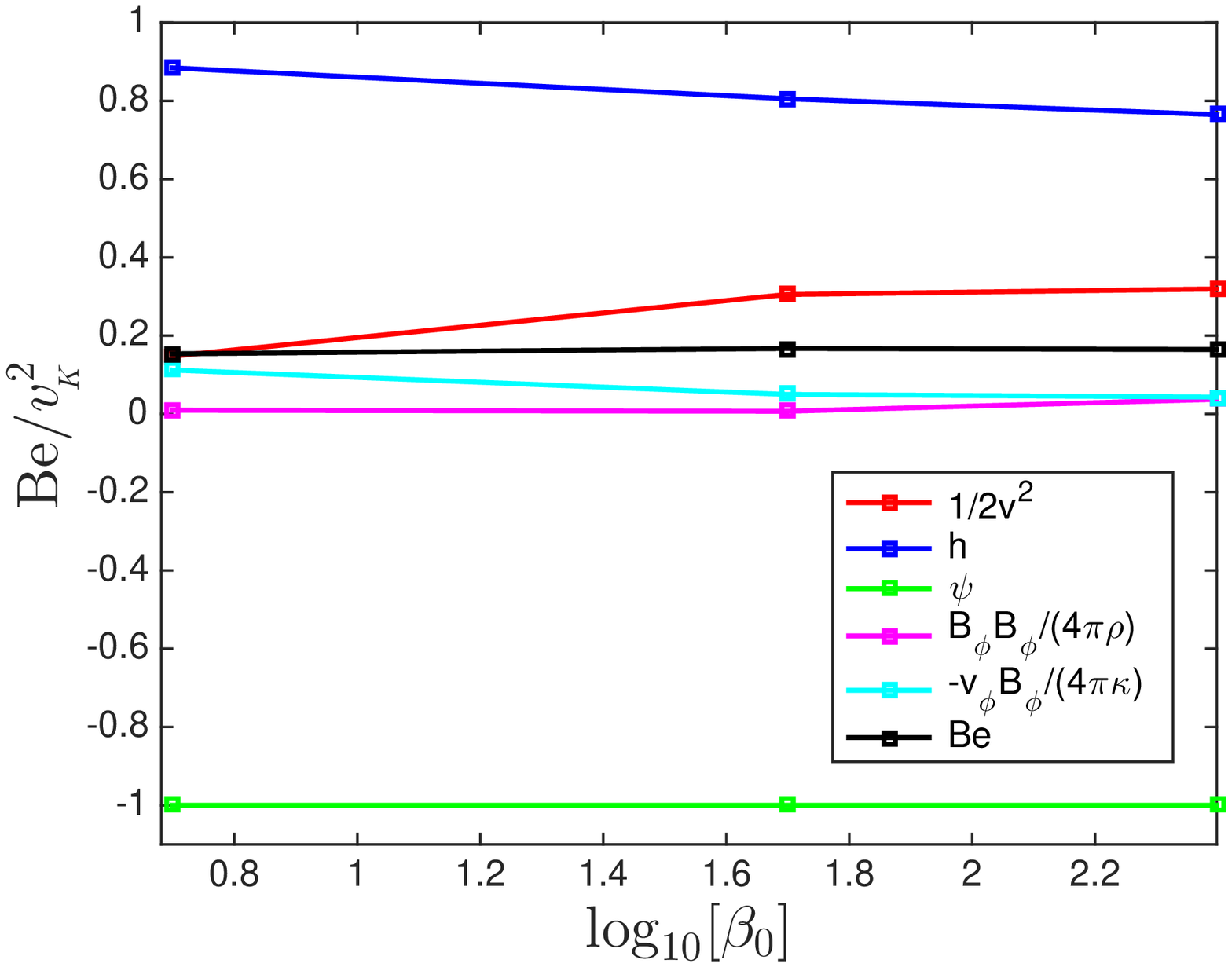}
\caption{Mass flux weighted Bernoulli parameter in unit of $ v_{\rm K}^2 $. The squares denote results in models A, B and C. \label{Bernoulli}}
\end{center}
\end{figure*}

From Figure \ref{velocites}, we can see that from the equatorial plane to the rotation axis, the density and pressure decrease. From the bottom right panel of Figure \ref{velocites}, we see that the temperature (or sound speed) in model C is slightly higher than those in models A and B. The temperature of accretion flow depends on heating. In this paper, we have both viscous and magnetic dissipation heating. Viscous heating $ q_{\rm vis} \propto (v_{\phi}/r)^{2} $. Magnetic dissipation heating $ q_{\rm res} \propto B^{2} $. In model C, the azimuthal velocity is much smaller than those in models A and B. Therefore, the viscous heating in model C is much smaller than those in models A and B. However, the magnetic dissipation rate in model C is much bigger than those in models A and B because magnetic field is much stronger in model C. The total heating rate per particle in model C is much bigger than those in models A and B. Quantitatively, we find than the total heating rate in model C is larger than those in models A and B by a factor of $ \sim 1.5 $. Therefore, the temperature of gas in model C is highest.

Bernoulli parameter is usually used to judge whether wind can escape to infinity. In magnetized accretion flow, Bernoulli parameter is defined as follows (\citealt{Zhu and Stone 2017}; \citealt{Fukue 1990}),
\begin{equation} \label{bernoulli parameter}
	Be = \frac{1}{2} v^{2} + h + \psi + \frac{B_{\phi} B_{\phi}}{4 \pi \rho} - \frac{B_{\phi} v_{\phi}}{4 \pi \kappa}
\end{equation}

Here, $ h = \gamma p/(\gamma-1)\rho $ is enthalpy, $\kappa=\rho v_p/B_p$ with $B_p=\sqrt{B_r^2+B_\theta^2}$. Figure \ref{Bernoulli} shows the mass flux weighted Bernoulli parameter. In all the models, the Bernoulli parameter is positive. Therefore, wind can escape to infinity. The enthalpy dominates other terms. The enthalpy in model C is just slightly bigger than those in models A and B. This is because, temperature (or sound speed) in model C is just slightly larger than those in models A and B (see bottom right panel of Figure \ref{velocites}). Due to the smaller azimuthal velocity in model C, the kinetic term in model C is smaller than those in models A and B by a factor of $ \sim 2 $. The magnetic term in model C is slightly larger than those in models A and B. However, the magnetic term is much smaller than other terms. The Bernoulli parameter almost does not change with the change of magnetic field strength. The mass flux-weighted Bernoulli parameter of the wind is evaluated as
\begin{equation} \label{Bernoulli1}
Be (r) = 0.16\ v_{\rm K}^{2}(r).
\end{equation}
Numerical simulation result of SANE model obtained by \citep[eq. 19]{Yuan et al. 2012b} showed that $ Be(r)/v_{\rm K}^2(r) \approx (0.1-0.2) $. Therefore, our present result is in good agreement with that obtained by numerical simulations.

\begin{figure*}
\begin{center}
\includegraphics[width=120mm, angle=0]{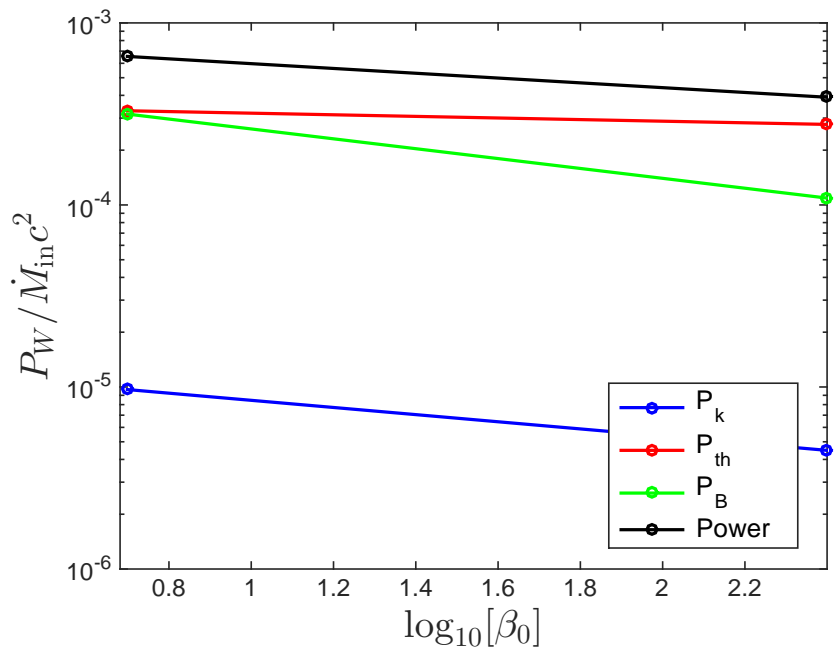}
\caption{Power carried by wind in unit of $ \dot M_{\rm in} c^2 $. The red and blue lines correspond to thermal and kinetic powers carried by wind, respectively. The green line shows Poynting flux. The black line shows the total power. \label{Power}}
\end{center}
\end{figure*}

We also calculate the kinetic and thermal energy carried by the wind as follows:

\begin{equation}
P_{\rm k} (r) = 2 \pi r^2 \int_{\rm 0^\circ}^{\rm 90^\circ} \rho \max(v_r^3,0) \sin\theta d\theta,
\end{equation}
\begin{equation}
P_{\rm th} (r) = 4 \pi r^2 \int_{\rm 0^\circ}^{\rm 90^\circ} \rho e \max(v_r,0) \sin\theta d\theta.
\end{equation}

The Poynting energy flux is
\begin{equation}
P_{\rm B} (r) = 4 \pi r^2 \int_{\rm 0^\circ}^{\rm 90^\circ} S_r \max(v_r/|v_r\emph{}|,0) \sin\theta d\theta.
\end{equation}

The radial component of Poynting flux $S_r$ is defined as (\citealt{Igumenshchev 2008}):
\begin{equation}
	S_r = v_r \frac{B^2}{4\pi} - \frac{B_r}{4\pi}(\bm{v} \cdot \bm{B})
\end{equation}

In this paper, the total power of wind is calculated as:
\begin{equation}
P_W = P_{\rm k} + P_{\rm th} + P_{\rm B}.
\end{equation}

We assume that the Poynting flux is part of the energy flux of wind. The reason is that it is believed finally the Poynting flux will be converted into kinetic power of wind (\citealt{Spruit 2010}; \citealt{Li and Cao 2010}). Figure \ref{Power} plots power carried by the wind. We have calculated the mass flux of wind in models A, B and C. We find that in model C, the mass flux of wind is just slightly increased by a factor of $ 0.03 $ compared to that in model A. The density and velocity of wind (see Figure \ref{velocites}) do not change much from models A to C. Therefore, the mass flux of wind in the three models do not differ much. The thermal power in model C is slightly larger than those in models A and B. This is because temperature and wind mass flux in model C are larger than those in models A and B. The Poynting flux in model C is $ 3 $ times that in model A. The reason is that magnetic field in model C is much larger than that in model A. Finally, the kinetic power also increases by a factor of $ 2 $ from model A to model C. The total power in model C is $ 1.6 $ times that in model A.

%-----------------------------------------------------------------
\section{Summary and discussion} \label{sec:summery}

In this paper, we solve two-dimensional MHD equations of hot accretion flow incorporating an evenly symmetric magnetic field. In order to simplify the equations, we assume radial self-similarity. We find that wind is present at $ \theta < 40^\circ $, while inflow is present around the equatorial plane. This structure is same as that obtained in previous numerical simulation works (e.g., \citealt{Yuan et al. 2012b}; \citealt{Narayan et al. 2012}; \citealt{Sadowski et al. 2013}; \citealt{Yuan et al. 2015}).

We focus on how the wind properties change with changing of magnetic field strength. We find that when the magnetic pressure is more than $ 2 $ orders of magnitude smaller than the gas pressure ($ \beta > 100 $; weakly magnetized flow), wind is driven by the combination of the gas pressure gradient and the centrifugal forces. However, when the magnetic field is strong ($\beta < 10$), wind is driven by the combination of the gas pressure gradient and the magnetic pressure gradient forces. We find that the Bernoulli parameter does not change much with the change of the magnetic field strength. The Bernoulli parameter of wind from both weakly and strongly magnetized accretion flow is about $ Be \sim 0.16 v_{\rm K}^2$.  We have also calculated the power carried by wind. We find that the power of wind in strongly magnetized flow is around $ 1.6 $ times that of wind in weakly magnetized flow.

The power of wind is in the range $10^{-4}-10^{-3} \dot M_{\rm in} c^2$. If wind is generated close to the black hole, then $ \dot M_{\rm in} $ is roughly equal to the accretion rate onto the black hole $ \dot M_{\rm BH} $. In numerical simulations studying evolution of galaxies (e.g., \citealt{Ciotti et al. 2010}; \citealt{Ostriker et al. 2010}; \citealt{Gaspari et al. 2012}), it is found that AGN wind can interact with the intercluster medium and heat the intercluster medium. The heating by AGN wind is useful to prevent rapid cool of the gas (i.e., the cooling flow problem). In order to be consistent with observations, the power of AGN wind should be in the range $ P_W \sim 10^{-4}-10^{-3} \dot M_{\rm BH} c^2 $. Thus, wind generated by hot accretion flow may play a role in solving the rapid cooling problem of intercluster medium when the AGN is in hot accretion mode.

%-----------------------------------------------------------------

\begin{acknowledgements}
We thank the referee for his/her thoughtful and constructive comments on an early version of the paper. We also thank Shuang-Liang Li and Zhao-Ming Gan for useful discussions.This work is supported in part by the National Program on Key Research and Development Project of China (Grant No. 2016YFA0400704),  the Natural Science Foundation of China (grants 11573051, 11633006, 11773053 and 11661161012), the Natural Science Foundation of Shanghai (grant 16ZR1442200), and the Key Research Program of Frontier Sciences of CAS (No. QYZDJSSW-SYS008). Amin Mosallanezhad is supported by Chinese Academy of Sciences President's International Fellowship Initiative, (PIFI). Grant No. 2018PM0046.  This work made use of the High Performance Computing Resource in the Core Facility for Advanced Research Computing at Shanghai Astronomical Observatory.
\end{acknowledgements}


\end{document}